The properties of FR0 radio galaxies as intermediate objects in the evolution of radio galaxies

David Garofalo[1], Chandra B. Singh[2], Eddie Harmon[1], Michael Williams[1], Luis Rojas Castillo[1],

1. Department of Physics, Kennesaw State University, USA
2. South-Western Institute for Astronomy Research (SWIFAR), Yunnan University, University Town, Chenggong, Kunming 650500, People's Republic of China;
chandrasingh@ynu.edu.cn

Abstract

It is becoming increasingly clear that counter-rotation between black holes and accretion disks is key to understanding radio galaxies. Such an accretion configuration was introduced over a decade ago to elucidate the nature of the radio loud/radio quiet dichotomy and the jet-disk connection, but has since been applied to a plethora of observations across space and time, from isolated to rich environments and from the formation of the first quasars to the nature of mature objects like M87. We briefly review the paradigm in which counter-rotation is key for the triggering of radio galaxies and its observational support, to then apply it to a series of seemingly unrelated observations concerning FR0 radio galaxies which we argue trace their explanation to one overarching theoretical idea. FR0 radio galaxies appear to be radio galaxies in transition, with low spinning black holes and thus weaker but tilted jets with respect to an earlier radio quasar phase. As a result of this jet reorientation, FR0 radio galaxies are prescribed to be in an earlier phase of star formation suppression in radio galaxies, compared to a later phase that is unlikely to be less than tens of millions of years in their future if they have enough accretion fuel to evolve into more powerful FRI radio galaxies. FR0 radio galaxies will have a greater or lesser star formation suppression feedback effect depending on how long they live. Tilted jets also enhance stellar velocities in the bulge. Because FR0 jet lengths are of the same order of magnitude as the radius of the stellar bulge, FR0 jets are prescribed to have begun, more or less recently depending on their age, to affect stellar velocity dispersions as well. As a result, they will be associated with dispersion values that tend to be larger than for characteristically non-jetted active galaxies, but smaller than for giant radio galaxies such as M87 that have experienced a long-term tilted and more powerful FRI jet. With these ideas it is possible to make a coarse-grained prediction for the slope on the M-σ plane for FR0 radio galaxies with values between 4 and 8.

1. Introduction

Powerful radio galaxies were classified 5 decades ago by Fanaroff & Riley, with FRII more powerful and collimated jets ( P > 2 x $10^{25}$ W/Hz at 178 Hz), often reaching 100 kpc before the environment claims their energy in hotspots, while FRI are less powerful and less collimated, more amenable to an interaction between jet and environment (Fanaroff & Riley 1974; Dabhade et al 2020; Rossi et al. 2020 and references therein). Although we will discuss a model with a precise time relation between jet kind, no consensus has emerged on the connection between FRII and FRI radio galaxies, if any. In fact, no agreement exists on the nature of FRI and FRII radio galaxies and why high excitation (HERG) and low excitation (LERG) radio galaxies can have FRII jets while FRI jets tend to be limited to low excitation (Prandoni et al 2007; Hardcastle et al 2007; Heckman & Best 2014; Garofalo, Singh & Zack 2018; Grandi et al 2021). Within the last decade and a half, a new class of LERG or low luminosity radio galaxy has been discovered, displaying compact jets that appear unresolved within 20 kpc of the galactic center. We argue from the perspective of our paradigm that their properties place them in-between powerful FRII and FRI radio galaxies. They are referred to as FR0 radio galaxies (Ghisellini et al 2011; Sadler et al. 2014; Baldi et al 2015). While it appeared that such radio galaxies might be the initial jet phases of either FRII or FRI radio galaxies (Baldi et al. 2018), evidence emerged that FR0s are not young radio galaxies, since their space density is larger than for objects they would evolve into (e.g. Capetti et al 2020), as the spin paradigm would claim (Lalakos et al. 2023), but interpreted as an intermediate phase in the evolution of radio galaxies in our paradigm given their mildly relativistic jets (Baldi et al. 2019).

Counter-rotation increasingly appears to be a key element in our understanding of jets from active galaxies (Raimundo et al 2023). Continued accretion spins counter-rotating black holes down and then up again with corotating accretion with FR0s fitting into the paradigm as transition objects, whose ancestors were powerful radio quasars (with counter-rotating accretion) that with sufficient cold gas evolve over time into FRI radio galaxies (with corotating accretion) (Garofalo & Singh 2019; Garofalo, Evans & Sambruna 2010). In-between such objects are radio galaxies in transition. If the accretion state around the black hole evolves away from high excitation and shifts to low excitation while still accreting in counter-rotation, it becomes an FRII LERG (Macconi et al 2020) whose properties have been shown to lie in-between radio quasars and FRI radio galaxies in terms of black hole mass and state of accretion (Garofalo & Singh 2020). But once the system is in corotation, no FRII jets are possible. The properties of FR0s (e.g. Giovannini et al 2023) allow us to fit them in the paradigm as low corotating black holes, with black hole spin values between 0.1 and 0.2. Crucially, the transition from counter-rotation to corotation involves a tilt in the accretion disk (i.e. the Bardeen-Petterson effect) and therefore in the jet direction (Garofalo, Joshi et al 2020). The result of this is a jet that sprays the interstellar medium face on, heating the environment, lowering star formation, and enhancing stellar velocity dispersion (Garofalo, Christian et al 2023). While this tilt in the disk and jet directions is attributable to the black hole engine, it allows the environment to become a crucial element in the coupling between jet and galaxy. Distinguishing engine-based features from environment-based features in jets is tricky. For example, a powerful jet in a dense medium will deposit its energy in a way that may be similar to the way a low power jet deposits its energy in less dense environments due to instabilities (e.g. Costa et al 2023). The timescale for this transition depends on the accretion rate which allows us to predict the environments where the tilt of the FR0 jet is visible. That FR0s are modeled as an early phase in the lifetime of FRI radio galaxies allows us to directly prescribe their location in the M-σ diagram. In Section 2 we review the so-called gap paradigm and describe the compatibility between observations and theory, make predictions, and prescribe in a coarse-grained way the location of FR0s in the M-σ plane. In Section 3 we conclude.

2. The gap paradigm and its application to FR0 radio galaxies

2.1 the gap paradigm

The gap paradigm (Garofalo, Evans & Sambruna 2010) for black hole accretion and jet formation is a scale-invariant model for accreting black holes that distinguishes itself from the spin-paradigm (Wilson & Colbert 1995; Sikora et al 2007) by considering counter-rotation between black hole and accretion disk triggered in galactic mergers and not in spiral galaxies, which instead, are triggered via secular processes. Because AGN triggered in spiral galaxies are limited to corotating black holes, jets, if formed, have limited power. The paradigm naturally incorporates an explanation to the fundamental unresolved issues in extragalactic astronomy of the 20[th] century, such as the jet-disk connection, the radio loud/radio quiet dichotomy, and the Owen-Ledlow diagram, including the mass dependence (Ghisellini & Celotti 2001). Because counter-rotation occurs under restrictive conditions (Garofalo, Christian & Jones 2019; King et al 2005), the fraction of non-jetted AGN is prescribed to dominate over the jetted AGN fraction in all environments and at any cosmic time. In its original formulation, the difference between FRII and FRI jets depends solely on black hole spin and disk orientation, with counter-rotation associated with FRII jets while corotation with FRI jets. This does not exclude jet power dependence on black hole mass. It has recently become clear through the so-called Roy Conjecture that jet morphology has a strong environmental component that is coupled to an engine-based component which is a tilt in the accretion disk in the transition through zero black hole spin (Garofalo, Moravec et al 2022). This development has allowed us to recently explain the two paths identified in the M-σ plane (Sahu et al 2019). For spinning black holes triggered in counter-rotation, accretion spins the black hole down toward zero spin and eventually into corotation. In the transition through zero black hole spin, the absence of frame-dragging and the Bardeen-Petterson effect (Bardeen & Petterson 1975) allow the incoming gas from the greater galaxy to form an accretion disk whose angular momentum retains that of the incoming gas. This means that a new plane of accretion is generated that is tilted with respect to the disk plane during counter-rotation (Garofalo, Joshi et al 2020). As a result of this tilt, when accretion has spun the black hole up to sufficiently large values in corotation to produce a jet, it will be pointing more directly into the thicker parts of the interstellar medium (ISM) where star formation is proceeding at an enhanced rate due in part to the earlier FRII jet (Garofalo, Christian et al 2016; Garofalo, Moravec et al 2022). Because of this direct coupling of jet with ISM (the Roy Conjecture), the jet will more readily deposit its energy in the ISM, thereby developing brighter regions closer to the galactic center, characteristic of FRI jets, unlike FRII jets which deposit that energy further out (Fanaroff & Riley 1974).

The tilt in the FRI jet and the timescale inferred from the model to go from the earlier FRII jet to the later FRI jet, allows the model to incorporate an explanation to the nature and origin of X-shaped radio galaxies and their preference for isolated environments (Garofalo, Joshi et al 2020). We showed that while jet reorientation occurs in rich environments as well, the timescale for evolution from FRII jet to FRI jet in clusters is typically orders of magnitude longer than in isolated environments because accretion rates in LERG systems are low, leaving no trace of the previous jet. Hence, no X-shaped radio galaxy. The possibility of a tilted jet that directly sprays its energy into the ISM has powerful consequences for the black hole scaling relations. The tilted jet, in fact, not only develops an FRI morphology, it also enhances stellar velocity dispersion in the bulge as well as eventually heating the ISM and lowering the star formation rate. This allows for a new understanding of the M-σ relations. We were able to show that the most powerful jetted and non-jetted AGN (e.g. radio loud and radio quiet quasars) populate the M-σ plane differently (Garofalo, Christian et al 2023). This tilted jet phenomenon allows for an association between powerful FRI radio galaxies, low rates of star formation, high stellar velocity dispersion, and bright nuclear stellar cores. This latest conclusion emerges from the origin of the radio galaxy as a radio quasar following the merger of the supermassive black hole binary in a major merger. Because counter-rotating accretion is more effective in merging the binary, the angular momentum of the binary can be extracted without appealing to the nuclear stellar core, leaving a bright stellar component in the nucleus of radio quasars. This will then characterize all the phases of the radio galaxy, including the latest and longest phase as an FRI radio galaxy. Because continuous accretion takes a counter-rotating system into a corotating one, the model allows us to make prescriptions for intermediate phases in radio galaxies. In the next section, we apply these ideas to the new subclass of FR0 radio galaxies.

2.2 FR0 radio galaxies

The general time evolution from FRII jets to FRI jets described in the previous section constitutes a general framework for understanding powerful jetted AGN, but it also offers a richer, subtler structure, that explains other subclasses of radio AGN. We now show how to understand FR0 radio galaxies within the framework outlined in Section 2.1. Capetti et al (2020) find most extended FR0 jets to reach 3-6 kpc but 18% of FR0CAT sources extend 10-15 kpc. According to Baldi (2023) FR0 jets reach as far out as 10-20 kpc. The difficulty in estimating length is associated with the difficulty in detecting faint sources. In Figure 1 we show the scale of radio contours for a typical FR0 radio galaxy from Baldi et al 2019. As a result of the fact that FR0 jets extend a few kpc, which is on the order of the galactic bulge (i.e. ~ 3 kpc scale ), we argue that such jets have at least begun to have an effect on stellar dispersion values. The possibility that they may remain compact jets for extended periods of time (for specific power and density values) , increases their feedback effect. We now show them to be transition objects in the evolution of radio galaxies.

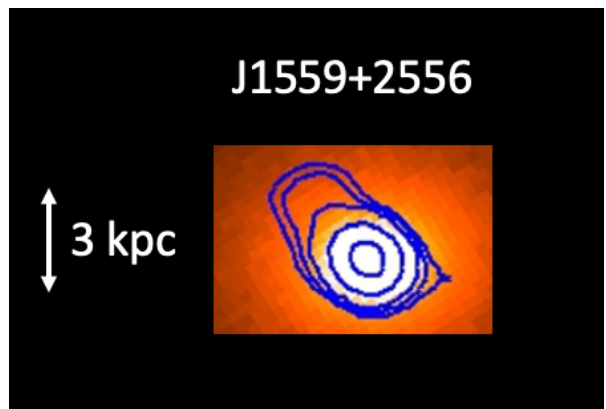

Figure 1: r-band SDSS image of the radio contours of an FR0 with scale (from Baldi et al 2019).

As mentioned, FR0 radio galaxies are an intermediate phase in the time evolution of originally merger-triggered, counter-rotating, radio quasars. In the paradigm, the latter are high-spinning black holes where misalignment between the angular momentum vector of the black hole and that of the incoming gas, ends up as a result of the Bardeen-Petterson effect (Bardeen & Petterson, 1975), in a thin disk that counter-rotates with respect to the black hole (Figure 2). The accretion disk is shown in blue and the jet in red. A high-spinning, counter-rotating black hole, spins down to zero spin in under $10^7$ years at the Eddington limit (Garofalo 2023). In denser environments, black holes tend to be more massive, and this translates into a stronger jet whose feedback alters the accretion mode from radiatively efficient and thin, to radiatively inefficient and thick. For the subclass of radio galaxies whose accretion mode changes in the transition through zero black hole spin in this way, an FR0 LERG emerges once the black hole spin crosses the 0.1 threshold and remains within this radio galaxy subclass until black hole spin crosses a value of about 0.2 (Garofalo & Singh 2019). At the Eddington limit, such a difference in spin is crossed in about $10^7$ years. However, since accretion has evolved into radiatively inefficient mode, the lifetime of the FR0 jet is enhanced. The lower the accretion rate, the longer the FR0 phase. In mergers where less gas is available, FR0 phases may be the end state of the radio galaxy because of insufficient gas to spin the black hole up beyond a spin value of 0.2, where the model prescribes that it will become an FRI radio galaxy (Garofalo & Singh 2019). In short, FR0 radio galaxies will tend to evolve more into FRI radio galaxies in the richest environments, where major mergers supply sufficient amount of accreting fuel. In less dense environments, such as groups, FR0 phases may last longer and in some cases never evolve beyond that before the active galaxy dies.

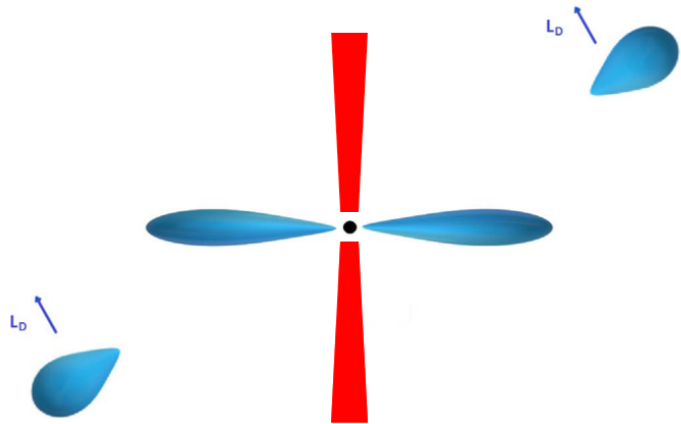

Figure 2: A jetted (red) spinning black hole. Frame dragging near the black hole forces a change in the angular momentum of the inflowing gas (blue), which forms a counter-rotating accretion disk (blue). The $L_D$ arrow represents the direction of the angular momentum of gas outside the black hole sphere of influence that will end up in the gravitational potential of the black hole to form the disk.

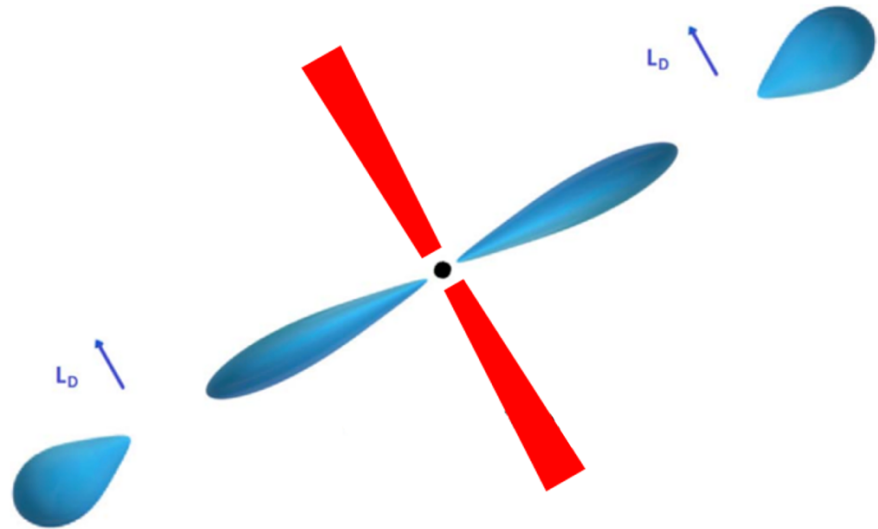

Figure 3: A corotating black hole that begins accreting when its spin is zero, shares the angular momentum of the inflowing cold gas from outside the black hole sphere of influence. The jet is therefore tilted with respect to the previous jetted phase
.

As shown by Garofalo, Joshi et al 2020, black holes that evolve from counter-rotation to corotation, will experience an absence of the Bardeen-Petterson effect, which allows the corotating FR0 jet to be tilted with respect to the previous counter-rotating jetted phase. This tilt in the jet (Figure 3) will eventually stifle star formation (the Roy Conjecture – Garofalo, Moravec et al 2022) but more readily enhance stellar velocity dispersions (Garofalo, Christian et al 2023). Because FR0 radio galaxies may evolve into the class of powerful FRI jetted active galaxies described in Garofalo, Christian et al 2023, FR0 radio galaxies must move on the M-σ plane in a way that situates them between the non-jetted AGN and the powerful jetted AGN. This idea can be made more quantitative.

For a given black hole mass, the FR0 phase is associated with lower jet power due to lower black hole spin ($0.1 < a < 0.2$). But the timescale for FR0 radio galaxies can vary dramatically, and this has implications for feedback. As described in Garofalo & Singh (2019), group environments which tend to have lower densities compared to cluster environments, will transition their counter-rotating accreting black holes into the corotating regime over longer timescales. At the Eddington accretion rate, the time to transition from corotating spin of 0.1 to corotating spin of 0.2 requires about $10^7$ years. Such objects are high excitation FR0 radio galaxies. Since the timescale between the relic FRII jet and the emerging FR0 jet is only a few million years, the FR0 jet is visible at the same time that the relic FRII jet is. But such objects are X-shaped radio galaxies and prescribed to emerge in isolated environments (see Figure 5 of Garofalo, Joshi et al 2020). For a complete understanding of the nature of FR0s, therefore, we must look more deeply into X-shaped radio galaxies.

Low excitation FR0 radio galaxies accrete at least two orders of magnitude below this value, which means such FR0s would live a billion years. In richer environments, the model prescribes accretion that eventually reach even lower rates such as in M87. In many FR0 radio galaxies with limited fuel, therefore, the FRI jetted phase will not be reached. The point is that the FR0 timescale or lifetime, $T_{FR0}$, spans a large range that is determined by the rate at which the black hole is fed, roughly

$$10^7 \text{ years} < T_{FR0} < 10^{10} \text{ years}.$$

Despite the relatively low power and limited size of FR0 jets, time can make a difference. If the timescale is sufficiently long, it will affect star formation rates in the galaxy at large, despite the compact size, and its effect on stellar dispersion will have also reached a more advanced phase. While such long-lasting FR0 radio galaxies are still part of an in-between phase of an evolutionary picture that leads to a more effective FRI radio galaxy feedback phase, it does not mean that FR0 phases are all equal in their feedback effect. For FR0 radio galaxies with timescales on the order of a few $10^8$ years, they would likely have some noticeable effect on stellar dispersion but a more mild one on the star formation rate, and would thus display bluer colors (Sadler et al 2014; Whittam et al 2016; Vardoulaki et al 2021). The timescale above for FR0 radio galaxies also explains why the relatively low observed abundance of high excitation FRII radio galaxies compared to FR0 radio galaxies does not preclude an evolutionary sequence between high excitation FRII radio galaxies and FR0 radio galaxies since the former live much shorter timescales. A high excitation FRII radio galaxy born with a rapidly spinning black hole accreting at the Eddington limit, in fact, will live about $8 \times 10^6$ years only.

Garofalo, Christian et al (2023) showed that separate M-σ paths exist for quasars with and without jets, with jetless quasars (e.g. radio quiet quasars) having a steeper M-σ slope due not only to absence of jets, but absence of tilted jets, which were shown to have a direct impact on star formation and stellar velocity dispersion. Because powerful radio quasars have an initial FRII phase associated with counter-rotation between black hole and disk, they will evolve through zero black hole spin and eventually develop tilted jets in the corotating spin regime (i.e. Figure 3). Since FR0s have jets that are extended enough to have an impact on the bulge, it follows that stellar dispersion is affected. Hence, a clear prediction is that FR0s must distribute themselves in a way that makes their slopes on the M-σ plane less than that for non-jetted quasars, because the latter have an even weaker feedback effect on stellar dispersion. The greater the interaction between FR0 jet and stellar bulge, the greater the effect on stellar dispersion. Hence, a simple prediction follows, which is that more extended and/or longer lasting FR0 jets will have larger stellar dispersion values. Despite the qualitative picture, we can make a clear quantitative prediction concerning the slope of FR0 radio galaxies on the M-σ plane, which is that they are between the large slopes of radio quiet quasars and the lower slope of radio loud quasars, which are characterized by

$$\log (M_{BH}/M_{solar}) = 8.64 \log (\sigma/200 \text{ km s}^{-1}) + 7.91$$

and

$$\log (M_{BH}/M_{solar}) = 4.95 \log (\sigma/200 \text{ km s}^{-1}) + 8.28,$$

respectively. The slope for non-jetted quasars is 8.64 while for jetted quasars it would be 4.95. We show this in Figure 4 with the predicted FR0 line in green. Note that these are the slopes of core Sersic and early type Sersic galaxies from Sahu et al (2019), which were connected to radio quiet and radio loud quasars in Garofalo, Christian et al (2023). As a result of the fact that counter-rotating accretion disks solve the last parsec problem without appealing to stars to rid the black hole binary of angular momentum (unlike corotating black holes), powerful radio galaxies will form with a full nuclear stellar component, making them Sersic galaxies. And FR0 radio galaxies are part of this family of AGN. The evolution of radio galaxies in our paradigm, and the transition from one type of jet to another, as well as a non-jetted phase as zero black hole spin is crossed, addresses the apparent conflicting evidence associating jets with different nuclear stellar densities (Balmaverde & Capetti 2006; Baldi et al 2010; Hamilton 2010; Richings et al 2011; Capetti et al 2023; Capetti & Brienza 2023; Garofalo, Christian et al 2023).

It also follows from theory that if you explore a sample of FR0 radio galaxies and compare it to FRI radio galaxies, one should find that while stellar velocity dispersion values occupy a range with higher values for the latter, the star formation rate will display a range of lower values for the latter. This is because tilted jets affect stellar dispersion and eventually also lower star formation rates, but FR0 radio galaxies have jets with a comparably weaker ability to suppress star formation so their star formation rates will have characteristically higher values than for FRIs which have much more extended jets (i.e. 100 kpc scale or more) as well as longer lifetimes. Figure 4 is not providing the range of values for black hole mass and stellar velocity dispersion, but the slope and intercept that will result from a fit to objects that have lower black hole mass values and lower stellar dispersion values compared to the most powerful radio galaxies (i.e. the red line in Figure 4).

In comparing FR0 radio galaxies to FRI radio galaxies, it is worth mentioning that selection effects exist. FR0s, being transition objects between radio galaxies, have lifetimes that depend on the accretion rate. A greater accretion rate will make the 0.1 to 0.2 black hole spin range shorter and therefore shorter FR0 lifetimes. As a result, FR0 radio galaxies will live longer in environments that are less dense, even by more than an order of magnitude. FR0s, in fact, prefer groups over clusters and have lower black hole masses than typical FRI radio galaxies. It is also worth emphasizing that FR0 radio galaxies with high excitation or FR0 HERGs are predicted to exist in environments where accretion rates have not dropped below $10^{-2}$ Eddington. But such conditions are prescribed to take place in isolated environments with smaller black hole masses and weaker jet feedback. These FR0 HERGs, again, come in the form of X-shaped radio galaxies (Garofalo, Joshi et al 2020) and are therefore theoretically distinguishable from small-scale jets in predominantly radio quiet Seyfert galaxies, by way of environment. Whereas this tilted jet phenomenon is prescribed to occur more in denser environments, it is also more difficult to detect in such environments because the transition through zero black hole spin is prescribed to often occur at low accretion rates in such environments, corresponding to longer timescales for reaching the prograde black hole spin values compatible with jets. It has been estimated that jet signatures last a few million years (Garofalo, Joshi et al 2020 and references therein), which is the timescale to bridge low counter-rotating black hole spin with low corotating black hole spin for black holes accreting at the Eddington limit. But in low excitation accretion systems, that timescale is increased by at least a factor of 100, bringing the timescale to at least 500 million years, which therefore washes out any signatures of a previous jet phase. This explains why current studies of parsec-scale and kpc-scale radio emission in FR0 radio galaxies do not find evidence of tilted jets (Cheng et al 2018; Cheng et al 2021; Capetti et al 2020; Baldi et al 2019; Baldi et al 2022; Giovannini et al 2023). For signatures of reorientation in rich environments see the very recent exploration of Ubertosi et al (2024) where VLBA data are compared with X-ray cavity images.

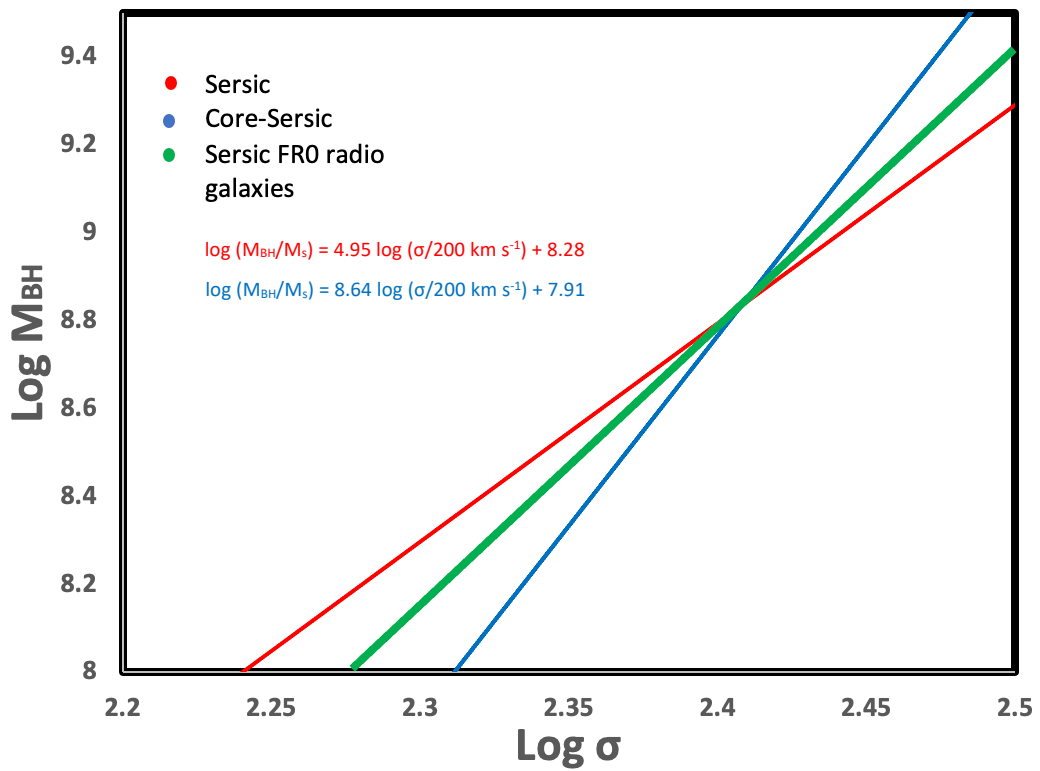

Figure 4: Prediction for the slope of FR0 radio galaxies in the M-σ plane sandwiched between the Sersic and core-Sersic galaxies of Sahu et al 2019. The green line simply means that our prediction requires FR0 radio galaxies to have a slope that lives in-between the red and blue lines.

3. Conclusions

We conclude with the following properties for FR0 radio galaxies.

- The black hole of an FRII radio quasar triggered in a merger must spin down toward zero spin and continued accretion will spin it up in corotation between disk and black hole. If accretion rates drop sufficiently, the end state of this continued evolution is an FRI radio galaxy. Because FR0 radio galaxies are prescribed to have spin between 0.1 and 0.2 in corotation, they must be intermediate objects in the evolution of radio galaxies. On average, therefore, they will have black hole masses that tend to be larger than FRII radio galaxies but less massive than FRI radio galaxies.

- The space density of different subclasses of radio galaxies depends on the lifetime of the radio galaxy. FR0 radio galaxies exhibit lifetimes that are up to orders of magnitude longer than FRII radio galaxies. The timescale depends on the accretion rate which is often also in transition between near Eddington values and a few orders of magnitude lower than that. FRI radio galaxies have the longest characteristic lifetimes which are on the order of the age of the universe. While FR0 radio galaxies may form in isolated or rich environments, they will tend to come in the form of X-shaped radio galaxies in isolated environments because accretion rates tend on average to be near Eddington values. Because less gas is available in environments of intermediate richness, FR0 radio galaxies will tend to live longer in such environments.

- Accretion configurations that end up in counter-rotation around the merged black holes tend to more easily eliminate the binary angular momentum without appealing to the stellar component, which implies that objects that were originally radio quasars, should have a brighter stellar nucleus compared to radio quiet quasars that never experience a counter-rotating phase. That FR0 radio galaxies are part of the sequence of objects connected to radio quasars implies that they too will have a bright stellar nucleus.

- Because FR0 radio galaxies tend to have tilted jets, they affect stellar velocity dispersion in their bulge and star formation by heating the gas. These properties are less dramatic than in fully developed FRI radio galaxies, which allows us to prescribe their location on the M-σ plane. The data on star formation, stellar dispersion, and nuclear stellar density, from JWST and LEGA-C, among others, should already be available for analysis.

- The tilted jet phenomenon that occurs in the transition through zero black hole spin allows us to string together a large body of seemingly unrelated observations of galaxies across space and time and ultimately supports the idea that counter-rotation in accretion is a key element in understanding the active phase of a galaxy.

Acknowledgments

CBS is supported by the National Natural Science Foundation of China under grant No. 12073021. We thank the anonymous referee for crucial insights into aspects of our paper that required explanation and further depth.